\documentclass [6pt,a4paper]{article}
\usepackage{bbm}
\usepackage{amsfonts}
\usepackage{mathrsfs}
\usepackage {amssymb}
\usepackage {amsmath}
\usepackage{amsthm}
\usepackage{latexsym}
\usepackage{indentfirst}
\usepackage{enumitem}

\newcommand{\abs}[1]{\left|#1\right|}

\newcommand{\ceil}[1]{\left\lceil#1\right\rceil}

\def\dse#1{\vskip 0.6cm\noindent
        {\large\bf #1}
        \vskip 0.4cm}
\def\dse#1{\vskip 0.6cm\noindent
        {\large\bf #1}
        \vskip 0.4cm}
\usepackage{lastpage}
\usepackage{epsfig}

\usepackage{anysize}
\marginsize{3cm}{3cm}{2cm}{2cm}
\normalsize

\begin{document}
\begin{center}
\textbf{\large{The MacWilliams identity for $m$-spotty weight
enumerator over
$\mathbb{F}_2+u\mathbb{F}_2+\cdots+u^{m-1}\mathbb{F}_2$}}\footnote {
This research is partially supported by NNSF of China (61202068,
11126174), Talents youth Fund of Anhui Province Universities
(2012SQRL020ZD). E-mail address: smjwcl.good@163.com (Min-Jia
Shi).}
\end{center}

\begin{center}
MinJia Shi
\end{center}

\begin{center}
\textit{\footnotesize School of Mathematical Sciences of Anhui
University, $230601$ Hefei, Anhui, China\\}
\end{center}

\noindent\textbf{Abstract}  Past few years have seen an extensive
use of RAM chips with wide I/O data (e.g. 16, 32, 64 bits) in
computer memory systems. These chips are highly vulnerable to a
special type of byte error, called an $m$-spotty byte error, which
can be effectively detected or corrected using byte error-control
codes. The MacWilliams identity provides the relationship between
the weight distribution of a code and that of its dual. The main
purpose of this paper is to present a version of the MacWilliams
identity for $m$-spotty weight enumerators over
$\mathbbm{F}_{2}+u\mathbbm{F}_{2}+\cdots+u^{m-1}\mathbbm{F}_{2}$
(shortly $R_{u, m, 2}$). 

\noindent\emph{keywords}: Byte error-control codes; $m$-spotty byte error; MacWilliams identity

\dse{1~~Introduction}

The error control codes have a significant role in improving
reliability in communications and computer memory system [1]. For
the past few years, there has been an increased usage of
high-density RAM chips with wide I/O data, called a byte, in
computer memory systems. These chips are highly vulnerable to
multiple random bit errors when exposed to strong electromagnetic
waves, radio-active particles or high-energy cosmic rays. To
overcome this, a new type of byte error known as spotty byte error
has been introduced in which the error occurs at random $t$ or fewer
bits within a $b$-bit byte [13], if more than one spotty byte error
occur within a $b$-bit byte, then it is known as a multiple spotty
byte error or $m$-spotty byte error [12]. To determine the
error-detecting and error-correcting capabilities of a code, some
special types of polynomials, called weight enumerators, are
studied.

One of the most celebrated results in the coding theory is the
MacWilliams identity that describes how the weight enumerator of a
linear code and the weight enumerator of the dual code relate to
each other. This identity has found widespread application in the
coding theory [2]. Recently various weight enumerators with respect
to $m$-spotty weight have been introduced and studied. Suzuki et al.
[12] defined Hamming weight enumerator for binary byte error-control
codes, and proved a MacWilliams identity for it. M. \"{O}zen and V.
Siap [3] and I. Siap [9] extended this result to arbitrary finite
fields and to the ring $\mathbb{F}_2+u\mathbb{F}_2$ with $u^2=0$,
respectively. I. Siap [10] defined $m$-spotty Lee weight and
$m$-spotty Lee weight enumerator of byte error-control codes over
$Z_4$ and derived a MacWilliams identity. A. Sharma and A. K. Sharma
introduced joint $m$-spotty weight enumerators of two byte
error-control codes over the ring of integers modulo $l$ and over
arbitrary finite fields with respect to $m$-spotty Hamming weight
[7], $m$-spotty Lee weight [5] and $r$-fold joint $m$-spotty weight
[6]. M. \"{O}zen and V. Siap [4] proved a MacWilliams identity for
the $m$-spotty RT weight enumerators of binary codes, which was
generalized to the case of arbitrary finite fields and ring $Z_l$ by
M. J. Shi [8]. In this paper, we will consider the MacWilliams
identity for $m$-spotty weight enumerators of linear codes over
$R_{u, m, 2}$, which generalizes the result of [9]. The organization
of this paper is as follows: Section 2 provides definitions of
$m$-spotty weight and some basic knowledge about $R_{u, m, 2}$.
Section 3 presents the MacWilliams identity for $m$-spotty weight,
and Section 4 illustrates the weight distribution of the $m$-spotty
byte error control code with an example.

\dse{2~~Preliminaries}

Consider the finite commutative ring $R_{u, m,
2}=\mathbbm{F}_{2}[u]/\langle
u^m\rangle=\mathbbm{F}_{2}+u\mathbbm{F}_{2}+\cdots+u^{m-1}\mathbbm{F}_{2}$
with $u^m=0$, when $m=1$, the ring $\mathbbm{F}_{2}$ is a field,
when $m=2$, the ring $R_{u, 2, 2}=\mathbb{F}_2+u\mathbb{F}_2$. In
the rest of this paper, we assume that $m\geq3$ is a positive
integer. Any linear code $C$ over $R_{u, m, 2}$ is permutation
equivalent to a code with generator matrix:
\begin{eqnarray}
\label{generator-C} G_1=\left(
\begin{array}{cccccccccccccccc}
I_{k_1} & A_{11}   & A_{12}       & A_{13}       &\cdots & A_{1,{m-1}}        &  A_{1m}\\
0       & uI_{k_2} & uA_{22}      & uA_{23}      &\cdots & uA_{2, m-1}        &  uA_{2, m}\\
0       & 0        & u^{2}I_{k_3} & u^{2}A_{33}  &\cdots & u^{2}A_{3, m-1}    & u^{2}A_{3, m}\\
\cdot   & \cdot    & \cdot        & \cdot        &\cdots & \cdot              & \cdot\\
0       & 0        & 0            & 0            &\cdots & u^{m-1}I_{k_m}     & u^{m-1}A_{k, m}\\
\end{array} \right),\nonumber
\end{eqnarray}
where $I_{k_i}$ is $k_i\times k_i$ identity matrix, $k_i$ is a
nonnegative integer, and the columns are grouped into blocks of
length $k_1$, $k_2$, $\cdots$, $k_m$, $n-k$ where
$k=\sum\limits_{i=1}^{m}{k_i}$, $k$ being the number of rows of
$G_1$. Any code $C$ has a generator matrix in the standard form, and
the parameters $k_1$, $k_2$, $\cdots$, $k_m$ are the same for any
generator matrix $G_1$ in the standard form for $C$. Each codeword
of $C$ can be expressed in the form $(x_1, x_2, \cdots, x_m)G_1$
where each $x_i$ is a vector of length $k_i$ with components in
$R_{u, m-i+1, p}$. Thus, $C$ has $2^{s}$ codewords where
$s=\sum\limits_{i=1}^{m}{(m-i+1)k_i}$.\\

\textbf{\upshape Definition 2.1} (see [13]). A spotty byte error is
defined as $t$ or fewer bits errors within a $b$-bit byte, where $1
\leq t \leq b$. When none of the bits in a byte is in error, we say
that no spotty byte error has occurred.\\

We can define the $m$-spotty weight and the $m$-spotty distance over
$R_{u, m, 2}$ as follows.

\textbf{\upshape Definition 2.2}. Let $e\in R_{u, m, 2}$ be an error
vector and $e_i\in R_{u, m, 2}^b$ be the $i$-th byte of $e$, where
$N=nb$ and $1 \leq i \leq n$. The number of $t/b$-errors in $e$,
denoted by $w_{M}(e)$, and called $m$-spotty weight is defined as
$$w_{M}(e)=\sum\limits_{i=1}^{n} \Big\lceil \frac{w_{M}(e_i)}{t} \Big\rceil,$$
where $\lceil x \rceil$ denotes the smallest integer not less than
$x$. If $t=1$, this weight, defined by $w_{M}$, is equal to the
Hamming weight. In a similar way, we define the $m$-spotty distance
of two codewords $u$ and $v$ as $d_M=\sum\limits_{i=0}^{n}
\Big\lceil \frac{d_H(u_i, v_i)}{t} \Big\rceil$. Further, it is also
straightforward to show that this distance is a metric in $R_{u, m,
2}^N$.

Hereinafter, codes will be taken to be of length $N$ where $N$ is a
multiple of byte length $b$, i.e. $N=nb$.

Let $c=(c_1, c_2, \cdots, c_N)$ and $v=(v_1, v_2, \cdots, v_N)$ be
two elements of $R_{u, m, 2}^N$. The inner product of $c$ and $v$,
denoted by $\langle c, v\rangle$, is defined as follows:
$$\langle c, v\rangle=\sum\limits_{i=1}^{n}\langle c_i, v_i\rangle=\sum\limits_{i=1}^{n}\Big(\sum\limits_{j=1}^{b}c_{i, j}v_{i,j}\Big).$$
Here, $\langle c_i, v_i\rangle=\sum\limits_{j=1}^{b}c_{i, j}v_{i,
j}$ denotes the inner product of $c_i$ and $v_i$, respectively.

Let $C$ be a linear code over $R_{u, m, 2}^N$. The set
$C^\bot=\{v\in R_{u, m, 2}^N|\langle u, v\rangle=0, {\rm for\ all}\
u\in C\}$ is also a linear code over $R_{u, m, 2}$ and it is called
the dual code of $C$.

\dse{3~~The MacWilliams identity}

Every element $x\in R_{u, m, 2}$ can be written uniquely as
$x=r_0(x)+ur_1(x)+\cdots+u^{m-1}r_{m-1}(x)$, where $r_i(x)\in
\mathbbm{F}_2, i=0, 1, \cdots, m-1$. It is easy to check that $x$ is
a unit if and only if $r_0(x)=1$. Hence there are $$\binom {m-1}
1+\binom {m-1} 2+\cdots+\binom {m-1} {m-1}=2^{m-1}$$ units in $R_{u,
m, 2}^{*}=R_{u, m, 2}\setminus \{0\}$. The number of zero divisors
in $R_{u, m, 2}^{*}$ is equal to
$2^{m-1}-1$.\\

In $R_{u, m,
2}$, there exists the chain of ideals: \\
$$R_{u, m,
2}=\langle1\rangle \supset \langle u\rangle\supset \langle
u^2\rangle\supset \cdots \supset \langle u^{m-1}\rangle \supset
<u^{m}>=<0>.$$

\textbf{\upshape Definition 3.1}. Define sets $A$ and $B$ as
follows: we first divide the elements of $R_{u, m, 2}$ into two
equal parts such that
\begin{itemize}\addtolength{\itemsep}{-5pt}
\item[(i)]$0$ and $1$ always belong to the set $A$;
\item[(ii)] both parts have the same number of zero divisors (Here, we
consider element 0 as a `` zero divisior") and they split each ideal
as well as all the units;
\item[(iii)] For all $a, b\in A$, then $a+b\in A$, i.e. the set $A$ is
closed with respect to addition;
\item[(iv)] For all $a, b\in B$, then $a+b\in A$;
\item[(v)] For all $a\in A, b\in B$, then $a+b\in B$.
\end{itemize}
Moreover, sets $A$ and $B$ are uniquely determined in the above
Definition.\\

 \textbf{\upshape Example 3.1.} Consider the ring $R_{u, 4, 2}=\mathbb{F}_2+u\mathbb{F}_2+u^2\mathbb{F}_2+u^3\mathbb{F}_2$. In $R_{u, 4, 2}$,
there exists the chain of ideals: $R_{u, 4, 2}=\langle1\rangle
\supset \langle u\rangle\supset \langle u^2\rangle \supset \langle
u^{3}\rangle=\langle0\rangle,$ where $\langle u\rangle=\{0, u, u^2,
u^3, u+u^2, u+u^3, u^2+u^3, u+u^2+u^3\}$, $\langle u^2\rangle=\{0,
u^2, u^3, u^2+u^3\}$ and $\langle u^3\rangle=\{0, u^3\}$. Hence,
according to Definition 3.1, we can obtain $$A=\{0, 1, u, 1+u, u^2,
u+u^2, 1+u^2, 1+u+u^2\}$$ and
$$B=\{u^3, u+u^3, u^2+u^3, u+u^2+u^3, 1+u^3, 1+u+u^3, 1+u^2+u^3,
1+u+u^2+u^3\}.$$

\textbf{\upshape Definition 3.2}. We define the character $\chi$:
\begin{equation}
\chi(a)=
\begin{cases}
1, & \text{if $a\in A$;} \\
-1, & \text{if } a\in B,
\end{cases}
\end{equation}
where sets $A$ and $B$ are defined as in Definition 3.1. We note
that $\chi$ is a nontrivial character, i.e. $\chi$ is not the
identity map on the nonzero ideals of $R_{u, m, 2}$. Moreover, for
all $a,
b\in R_{u, m, 2}$, we have $\chi(a+b)=\chi(a)\cdot\chi(b)$.\\

\textbf{\upshape Definition 3.3}. Let $v=(v_1, v_2, \cdots, v_b)\in
R^b$. Then the support of $v$ is defined by ${\rm
supp}(v)$$=\{i|v_i\neq0\}$
and the complement of ${\rm supp}(v)$ is denoted $\overline{{\rm supp}(v)}$.\\

\textbf{\upshape Definition 3.4}. Let $c=(c_1, c_2, \cdots, c_b)\in
R^b$ and define
\begin{align*}
S_k(c)&=\{v\in R^b| {\rm supp}(v)\subseteq {\rm supp}(c)\ {\rm and}
\ k=|{\rm
supp}(v)|\} \ {\rm and} \\
\overline{S}_k(c)&=\{v\in R^b| {\rm supp(v)}\subseteq \overline{\rm
supp(c)} \ {\rm and} \ k=|{\rm supp}(v)|\}.
\end{align*}

In order to prove our main theorem, we should first introduce the
following lemmas.

\textbf{\upshape Lemma 3.1}. Let $H\neq 0$ be an ideal of $R_{u, m,
2}$. Then
$$\sum_{a\in H}\chi(a)=0.$$

\textbf{{Proof}}. We can obtain the result readily by using the
definition of character $\chi$ in (1).\\

\textbf{\upshape Lemma 3.2}. Let $a\in R_{u, m, 2}$. Then
\begin{equation}
\sum_{r\in R_{u, m, 2}}\chi(ar)=
\begin{cases}
2^m, & \text{if $a=0$,} \\
0, & \text{if } a\neq 0.
\end{cases}
\end{equation}

\textbf{{Proof}}. If $a=0$, then clearly $\chi(ar)=1$ for all $r\in
R_{u, m, 2}$ and hence the result follows. Otherwise, if $a$ is a
unit, then elements $ar$, for all $r\in R_{u, m, 2}$, run over all
elements of $R_{u, m, 2}$, which forms a trivial ideal $R_{u, m,
2}$. If $a (\neq0)$ is a zero divisor, then elements $ar$, for all
$r\in R_{u, m, 2}$, form a proper ideal of $R_{u, m, 2}$. Hence,
according to Lemma 3.1, if $a\neq0$, we have $\sum_{r\in
R}\chi(ar)=0$.\\

\textbf{\upshape Lemma 3.3}. Let $v=(v_1, v_2, \cdots, v_b)\in R_{u,
m, 2}^b$, with $w(c)=j\neq 0$ and $k\in \{1, 2, \cdots, j\}$. Then
we have $$\sum\limits_{0\leq w(v)\leq k\atop {\rm supp}(v)\subseteq
{\rm supp}(c)}\chi(\langle c, v\rangle)=0.$$

\textbf{{Proof}}. The proof is similar to [9], so we omit it here.\\

\textbf{\upshape Lemma 3.4}. Let $c=(c_1, c_2, \cdots, c_b)\in R_{u,
m, 2}^b$ and $w(c)\neq0$. For all $k$ positive integers, we let
$I_k=\{i_1, i_2, \cdots, i_k\}\subseteq {\rm supp}(c)$ and
$I_0={\o}$. Then we have
\begin{equation}
\sum\limits_{v\in R_{u, m, 2}^b\atop {\rm supp}(v)=I_k}\chi(\langle
c, v\rangle)=(-1)^k.\nonumber
\end{equation}

\textbf{{Proof}}. For $k=0$ i.e. $I_0={\o}$, we have
$\sum\limits_{v\in R_{u, m, 2}^b\atop {\rm supp}(v)=I_0}\chi(\langle
c, v\rangle)=\sum\limits_{w(v)=0}\chi(0)=1$. For $k=1$, according to
Lemma 3.3, we have
\begin{equation}
\sum\limits_{v\in R_{u, m, 2}^b\atop {\rm supp}(v)=I_1}\chi(\langle
c, v\rangle)=\sum\limits_{i_1\in I_1\atop v_{i_1}\in R_{u, m,
2}^*}\chi(c_{i_1}v_{i_1})=\sum\limits_{v_{i_1}\in R_{u, m,
2}}\chi(c_{i_1}v_{i_1})-1=-1.\nonumber
\end{equation}
For $k=2$, according to Lemma 3.3, we have
\begin{eqnarray}
\sum\limits_{v\in R_{u, m, 2}^b\atop {\rm supp}(v)=I_2}\chi(\langle
c, v\rangle)&=&\sum\limits_{i_1, i_2\in I_2\atop v_{i_1}, v_{i_2}\in
R_{u, m, 2}^*}\chi(c_{i_1}v_{i_1}+c_{i_2}v_{i_2})\nonumber \\
\nonumber &=&\sum\limits_{i_1,i_2\in I_2\atop v_{i_1}, v_{i_2}\in
R_{u, m,
2}}\chi(c_{i_1}v_{i_1}+c_{i_2}v_{i_2})-\sum\limits_{i_1,i_2\in I_2
\atop v_{i_1}=v_{i_2}=0}\chi(c_{i_1}v_{i_1}+c_{i_2}v_{i_2})\nonumber
\\ \nonumber &&-\sum\limits_{v_{i_1}=0,\atop v_{i_2}\in R_{u, m,
2}*}\chi(c_{i_2}v_{i_2})-\sum\limits_{v_{i_2}=0,\atop v_{i_1}\in
R_{u, m, 2}*}\chi(c_{i_1}v_{i_1})\nonumber \\ \nonumber
&=&0-1-2(\sum\limits_{v_{i_1}\in R_{u, m,
2}}\chi(c_{i_1}v_{i_1})-1)=1.\nonumber \\ \nonumber
\end{eqnarray}
Now, we assume that the identity holds true for $k=r (r\geq3)$, i.e.
$\sum\limits_{v\in R_{u, m, 2}^b\atop {\rm supp}(v)=I_r}\chi(\langle
c, v\rangle)=(-1)^r$. For $k\geq r+1$, suppose ${\rm
supp}(v)$$=\{i_1, i_2, \cdots, i_{r+1}\}$. Then we have
\begin{eqnarray}
\sum\limits_{v\in R_{u, m, 2}^b\atop {\rm
supp}(v)=I_{r+1}}\chi(\langle c, v\rangle)
&=&\sum\limits_{i_1,i_2,\cdots,i_{r+1}\in I_{r+1}\atop v_{i_1},
v_{i_2}, \cdots, v_{i_{r+1}}\in
R_{u, m, 2}^*}\chi(\sum\limits_{j=1}^{r+1}c_{i_j}v_{i_j})\nonumber \\
\nonumber &=&\sum\limits_{i_1, i_2, \cdots, i_{r+1}\in I_{r+1}\atop
v_{i_1}, v_{i_2}, \cdots, v_{i_{r+1}}\in R_{u, m,
2}}\chi(\sum\limits_{j=1}^{{r+1}}c_{i_j}v_{i_j})-\binom {r+1}
1\sum\limits_{i_1, i_2, \cdots, i_{r}\in I_{r}\atop v_{i_1},
v_{i_2}, \cdots, v_{i_{r}}\in R_{u, m, 2}^*}\chi(\sum\limits_{j=1}^{r}c_{i_j}v_{i_j}) \nonumber \\
\nonumber &&-\binom {r+1} 2\sum\limits_{i_1, i_2,\cdots, i_{r-1}\in
I_{r-1}\atop v_{i_1}, v_{i_2}, \cdots, v_{i_{r-1}}\in R_{u, m,
2}^*}\chi(\sum\limits_{j=1}^{r-1}c_{i_j}v_{i_j})-\cdots\nonumber \\
\nonumber &&-\binom {r+1} {r}\sum\limits_{i_1\in I_{1}\atop
v_{i_1}\in R_{u,
m, 2}^*}\chi(c_{i_1}v_{i_1})-1\nonumber \\
\nonumber &=&0-\binom {r+1} 1(-1)^r-\binom {r+1}
2(-1)^{r-1}-\cdots-\binom
{r+1} {r-1}(-1)^{1}-1\nonumber \\
\nonumber &=&(-1)^{r+1}.
\end{eqnarray}
In the last line of equation above, we set $a=1, b=-1$ in the
following equation: $$(a+b)^{r+1}=\sum\limits_{i=1}^{r+1}\binom
{r+1} i a^ib^{r+1-i}.$$

\textbf{\upshape Corollary 3.1}. Let $c=(c_1, c_2, \cdots, c_b)\in
R_{u, m, 2}^b$ and $w(c)=j\neq 0$. For all $0\leq k\leq j$, we have
$$\sum\limits_{v\in S_k(c)}\chi(\langle c, v\rangle)=(-1)^k\binom j
k.$$

\textbf{{Proof}}. According to Definition 3.3 and Lemma 3.4, we get
\begin{equation}
\sum\limits_{v\in S_k{(c)}}\chi(\langle c,
v\rangle)=\sum\limits_{I_k\subseteq {\rm supp}(c)}\sum\limits_{{{\rm
supp}(v)=I_k}}\chi(\langle c, v\rangle)=\sum\limits_{I_k\subseteq
{\rm supp}(c)}(-1)^k=(-1)^k\binom j k.\nonumber
\end{equation}

\textbf{\upshape Lemma 3.5}. Let $c=(c_1, c_2, \cdots, c_b)\in R_{u,
m, 2}^b$ and $w(c)=j\neq0$. For all $0\leq k\leq j$, we have
\begin{equation}
\sum\limits_{v\in \overline{S}_k(c)}\chi(\langle c,
v\rangle)=(2^m-1)^k\binom {b-j} k.\nonumber
\end{equation}

\textbf{{Proof}}. Since $v\in \overline{S}_k(c)$ with ${\rm
supp}(v)\subseteq \overline{{\rm supp}(c)}$, we have $\chi(\langle
c, v\rangle)=1$. Further, since $k=|{\rm supp}(v)|$, there are
$\binom {b-j} k$ ways of choosing a subset of size $k$ from the
complement of support of $c$ of size $k$. For each subset of size
$k$, the sum of
characters equals to $(2^m-1)^k$. Hence, the result follows.\\

According to Lemma 3.5 and Corollary 3.1, we have the following
corollary.

 \textbf{\upshape Corollary 3.2}. Let $c=(c_1, c_2, \cdots,
c_b)\in R_{u, m, 2}^b$ and $w(c)=j$, $0\leq j_1\leq j$ and $0\leq
j_2\leq b-j$. We define $$S_{j_1, j_2}(c)=\big\{v\in R_{u, m,
2}^b|j_1=|{\rm supp}(v)\cap {\rm supp}(c)|\ {\rm and} \ j_2=|{\rm
supp}(v)\cap \overline{{\rm supp}(c)}|\big\}.$$ Then we can obtain
$$\sum\limits_{v\in s_{j_1, j_2}(c)}\chi(\langle c,
v\rangle)=(-1)^{j_1}(2^m-1)^{j_2}\binom j {j_1} \binom {b-j}
{j_2}.$$

The proof of the following two lemmas are similar to those of Lemma
2.7 and Lemma 2.8 in [9], so we omit it here.

\textbf{\upshape Lemma 3.6}. Let $c=(c_1, c_2, \cdots, c_b)\in R_{u,
m, 2}^b$ and $w(c)=j$. Then
$$\sum\limits_{v\in R^b}\chi(\langle
c, v\rangle)z^{\lceil
w_M(v)/t\rceil}=\sum\limits_{j_1=0}^{j}\sum\limits_{j_2=0}^{b-j}(-1)^{j_1}(2^m-1)^{j_2}\binom
j {j_1} \binom {b-j} {j_2}z^{\lceil(j_1+j_2)/t\rceil}.$$

The following lemma plays an important role in deriving the
MacWilliams identity for $m$-spotty weight.

\textbf{\upshape Lemma 3.7}. Let $C$ be a linear code of length $nb$
over $R_{u, m, 2}$ and $C^\bot$ its dual code and
$$\widehat{f}(u)=\sum\limits_{v\in R^{nb}}\chi(\langle
c, v\rangle)f(v).$$ Then $$\sum\limits_{v\in
C^\perp}=\frac{1}{|C|}\sum\limits_{u\in C}\widehat{f}(u).$$

Let $\alpha_j=\#\{i: w{(c_i)}=j, 1\leq i\leq n\}$. That is,
$\alpha_j$ is the number of bytes having Hamming weight $j$, $0\leq
j\leq b$, in a codeword. The summation of $\alpha_0, \alpha_1,
\cdots, \alpha_b$ is equal to the code length in bytes, that is
$\sum_{j=0}^{b}\lceil j/t \rceil=n$. The Hamming weight distribution
vector $(\alpha_0, \alpha_1, \cdots, \alpha_b)$ is determined
uniquely for the codeword $c$. Then, the $m$-spotty weight of the
codeword $c$ is expressed as $w_M{(c)}=\sum_{j=0}^{b}\lceil j/t
\rceil\cdot\alpha_j$. Let $A_{(\alpha_0, \alpha_1, \cdots, \alpha_b
)}$ be the number of codewords with Hamming weight distribution
vector $(\alpha_0, \alpha_1, \cdots, \alpha_b)$. For example, Let
$c=(0u0 \ u^20u^3\ 1uu^2\ 000\ u10)$ is a codeword with byte $b=3$.
Then the Hamming weight distribution vector of the codeword is
$(\alpha_0, \alpha_1, \alpha_2, \alpha_3)=(1, 1, 2, 1)$. Therefore,
$A_{(1, 1, 2, 1)}$ is the number of codewords with Hamming weight
distribution vector $(1, 1, 2, 1)$.

We are now ready to define the $m$-spotty weight enumerator of a
byte error control code over $R_{u, m, 2}$.

\textbf{\upshape Definition 3.5}. The weight enumerator for
$m$-spotty byte error control code $C$ is defined as
$$W(z)=\sum_{c \in C} z^{{w_M(c)}}.$$
By using the parameter $A_{(\alpha_0, \alpha_1, \cdots, \alpha_b)}$,
which denotes the number of codewords with Hamming weight
distribution vector $\{\alpha_0, \alpha_1, \cdots, \alpha_b\}$,
$W(z)$ can be expressed as follows:
\begin{eqnarray}
W(z)&=&\sum_{\substack{ (\alpha_0, \dots, \alpha_b) \\
\alpha_0, \dots, \alpha_b \ge 0 \\ \alpha_0 + \dots + \alpha_b = n }
} A_{(\alpha_0,\dots,\alpha_b)} \prod_{j=0}^b
(z^{\ceil{j/t}})^{\alpha_j}.\nonumber \\ \nonumber
\end{eqnarray}

\textbf{\upshape Theorem 3.1}. Let $C$ be a code over $R_{u, m, 2}$.
The relation between the $m$-spotty $t/b$-weight enumerators of $C$
and its dual is given by

\begin{eqnarray}
W^\perp (z) &=& \sum_{\substack{ (\alpha_0, \dots, \alpha_b) \\
\nonumber \alpha_0, \dots, \alpha_b \ge 0 \\ \alpha_0 + \dots +
\alpha_b = n } } A^\perp_{(\alpha_0,\dots,\alpha_b)} \prod_{j=0}^b
(z^{\ceil{j/t}})^{\alpha_j} = \frac{1}{\abs{C}}  \sum_{\substack{
(\alpha_0, \dots, \alpha_b)
\\ \alpha_0, \dots, \alpha_b \ge 0 \\ \alpha_0 + \dots + \alpha_b =
n } } A_{(\alpha_0,\dots,\alpha_b)} \prod_{j=0}^b (F_j^{(b,
m)}(z))^{\alpha_j},
\end{eqnarray}
where $F_j^{(b,
m)}(z)=\sum\limits_{j_1=0}^{j}\sum\limits_{j_2=0}^{b-j}(-1)^{j_1}(2^m-1)^{j_2}\binom
j {j_1} \binom {b-j} {j_2}z^{\lceil(j_1+j_2)/t\rceil}.$

\textbf{{Proof}}. Let $f(v)$ in Lemma 3.7 be considered as
$f(v)=z^{w_M(v)}$. Then the function $\tilde{f} (c)$ is calculated
as follows:
\begin{eqnarray}
\tilde{f} (c) &=& \sum_{v \in R_{u, m, 2}^{nb}} \chi(\langle c,
v\rangle) z^{w_M(v)} \nonumber
\\ \nonumber
&=& \sum_{v_1 \in R_{u, m, 2}^{b}}\sum_{v_2 \in R_{u, m,
2}^{b}}\cdots\sum_{v_n \in R_{u, m, 2}^{b}}\chi(\langle c_1,
v_1\rangle)\chi(\langle c_2, v_2\rangle)\cdots\chi(\langle c_n,
v_n\rangle)\prod_{i=1}^n z^{\ceil{w_H{(v_i)}/t}} \\ \nonumber &=&
\prod_{i=1}^n \left( \sum_{v_i \in R_{u, m, 2}^b} \chi(\langle c_i,
v_i\rangle) z^{\ceil{w_H{(v_i)}/t}} \right).
\end{eqnarray}
By applying Lemma 3.6, we have,
\begin{eqnarray}
\tilde{f}
(c)=\prod_{i=1}^n\left(\sum\limits_{j_1=0}^{k_i}\sum\limits_{j_2=0}^{b-{k_i}}(-1)^{j_1}(2^m-1)^{j_2}\binom
{k_i} {j_1} \binom {b-{k_i}}
{j_2}z^{\lceil(j_1+j_2)/t\rceil}\right),\nonumber
\end{eqnarray}
where $k_i=w(c_i)$. Thus
\begin{eqnarray}
\tilde{f}
(c)=\prod_{j=0}^b\left(\sum\limits_{j_1=0}^{j}\sum\limits_{j_2=0}^{b-j}(-1)^{j_1}(2^m-1)^{j_2}\binom
{j} {j_1} \binom {b-j}
{j_2}z^{\lceil(j_1+j_2)/t\rceil}\right)^{\alpha_j(c)},\nonumber
\end{eqnarray}
where $\alpha_j(c)=|\{i|w(c_i)=j\}|$.
\begin{eqnarray}
\tilde{f}
(c)=\prod_{i=1}^n\left(\sum\limits_{j_1=0}^{j}\sum\limits_{j_2=0}^{b-j}(-1)^{j_1}(2^m-1)^{j_2}\binom
{j} {j_1} \binom {b-j}
{j_2}z^{\lceil(j_1+j_2)/t\rceil}\right)^{\alpha_j(c)}.\nonumber
\end{eqnarray}
After rearranging the summations on both sides according to the
weight distribution vectors of codewords in $C^\perp$ and $C$
respectively, we have the result
\begin{eqnarray}
W^\perp (z) &=& \sum_{\substack{ (\alpha_0, \dots, \alpha_b) \\
\nonumber \alpha_0, \dots, \alpha_b \ge 0 \\ \alpha_0 + \dots +
\alpha_b = n } } A^\perp_{(\alpha_0,\dots,\alpha_b)} \prod_{j=0}^b
(z^{\ceil{j/t}})^{\alpha_j} = \frac{1}{\abs{C}}  \sum_{\substack{
(\alpha_0, \dots, \alpha_b)
\\ \alpha_0, \dots, \alpha_b \ge 0 \\ \alpha_0 + \dots + \alpha_b =
n } } A_{(\alpha_0,\dots,\alpha_b)} \prod_{j=0}^b (F^{(b,
m)}_j(z))^{\alpha_j}.
\end{eqnarray}

\dse{4~~Example}

Let \begin{equation} G=\left(\begin{array}{ccccccccc}
1 & 0 & 0   & u+u^2 & 0  & 0  \\
0 & u & 0   & u^2   & 0  & u^3\\
0 & 0 & u^2 & 0     & u^3& 0 \nonumber
\end{array}\right)
\end{equation}
be the generator matrix of a linear code $C$ over $R_{u, 4,
2}=\mathbb{F}_2+u\mathbb{F}_2+u^2\mathbb{F}_2+u^3\mathbb{F}_2$ of
length 6. $C$ has $2^9=512$ codewords. The dual code of $C$ is a
linear code of length 6 over $R_{u, 4, 2}$ and it has $2^{15}=32768$
codewords.

The Hamming weight distribution vectors of the codewords of $C$, the
number of codewords, and polynomials $F_j^{(b, m)}(z)$ for $b=3$ and
$t=2$ are shown in Tables 1 and 2 for the necessary computations to
apply the main theorem.

\begin{table}[thb]
\begin{center}
\small {\textbf{Table 1}\\ Hamming weight distribution vectors of
the
codewords\\ in $C$ and the number of codewords.}\\
{\begin{tabular}{cl}
\hline  $(\alpha_0, \alpha_1, \alpha_2, \alpha_3)$ \ \ \ \ \ \ \ \ \ \ \ & \ \ \ \ \ \ \ \ number \\
\hline $(2, 0, 0, 0)$\ \ \ \ \ \ \ \ \ \ \     & \ \ \ \ \ \ \ \ $1$\\
       $(0, 2, 0, 0)$\ \ \ \ \ \ \ \ \ \ \     & \ \ \ \ \ \ \ \ $18$\\
       $(0, 0, 2, 0)$\ \ \ \ \ \ \ \ \ \ \     & \ \ \ \ \ \ \ \ $88$\\
       $(0, 0, 0, 2)$\ \ \ \ \ \ \ \ \ \ \     & \ \ \ \ \ \ \ \ $104$\\
       $(1, 1, 0, 0)$\ \ \ \ \ \ \ \ \ \ \     & \ \ \ \ \ \ \ \ $3$ \\
       $(1, 0, 1, 0)$\ \ \ \ \ \ \ \ \ \ \     & \ \ \ \ \ \ \ \ $7$\\
       $(1, 0, 0, 1)$\ \ \ \ \ \ \ \ \ \ \     & \ \ \ \ \ \ \ \ $5$\\
       $(0, 1, 1, 0)$\ \ \ \ \ \ \ \ \ \ \     & \ \ \ \ \ \ \ \ $72$\\
       $(0, 1, 0, 1)$\ \ \ \ \ \ \ \ \ \ \     & \ \ \ \ \ \ \ \ $58$\\
       $(0, 0, 1, 1)$\ \ \ \ \ \ \ \ \ \ \     & \ \ \ \ \ \ \ \ $156$\\
\hline
\end{tabular}}
\\

\vskip 8pt

{\small \textbf{Table 2}\\ Polynomials $V_j^{(3, 4)}$ for $t=2$ and $b=3$}.\\
{\begin{tabular}{l}
\hline \ \ \ \ \ \ \ $F_0^{(3, 4)}(z)=1+720z+3375z^2$\ \ \ \ \ \ \ \\
       \ \ \ \ \ \ \ $F_1^{(3, 4)}(z)=1+224z-225z^2$\ \ \ \ \ \ \ \\
       \ \ \ \ \ \ \ $F_2^{(3, 4)}(z)=1-16z+15z^2$\ \ \ \ \ \ \ \\
       \ \ \ \ \ \ \ $F_3^{(3, 4)}(z)=1-z^2$\ \ \ \ \ \ \ \\
\hline
\end{tabular}}
\end{center}
\end{table}
According to the expression of $W(z)$ and Table 1, we obtain the
$m$-spotty weight enumerator of $C$ as
$$W(z)=1+10z+183z^2+214z^3+104z^6.$$
By applying Theorem 3.1 and Table 2, we obtain
\begin{eqnarray}
W^\perp (z)&=& \frac{1}{\abs{C}}  \sum_{\substack{  \alpha_0+
\alpha_1+\alpha_2+\alpha_3=2}} A_{(\alpha_0, \alpha_1, \alpha_2,
\alpha_3)} \prod_{j=0}^3 (F_j(z))^{\alpha_j}  \nonumber \\
\nonumber &=&\frac{1}{512}\big[(F_0^{(3, 4)}(z))^2+18(F_1^{(3,
4)}(z))^2+88(F_2^{(3, 4)}(z))^2+104(F_3^{(3, 4)}(z))^2+3F_0^{(3,
4)}(z)F_1^{(3, 4)}(z)\nonumber
\\ \nonumber
&&+7F_0^{(3, 4)}(z)F_2^{(3, 4)}(z)+5F_0^{(3, 4)}(z)F_3^{(3,
4)}(z)+72F_1^{(3, 4)}(z)F_2^{(3, 4)}(z)\nonumber
\\ \nonumber
&&+156F_2^{(3, 4)}(z)F_3^{(3, 4)}(z)+58F_1^{(3, 4)}(z)F_3^{(3, 4)}(z)\big]\nonumber \\
\nonumber &=&1+85z+3153z^2+9707z^3+19822z^4.
\end{eqnarray}
\dse{5~~Conclusion}

This paper has presented the MacWilliams identity for $m$-spotty
weight enumerators of the $m$-spotty byte error control codes. This
provides the relation between the $m$-spotty weight enumerator of
the code and that of the dual code. Also, the indicated identity
includes the MacWilliams identity over $\mathbb{F}_2+u\mathbb{F}_2$
in [9] as a special case.


\begin{thebibliography}{11}\addtolength{\itemsep}{-5pt}

\bibitem{pa0} Fujiwara  E.: Code design for dependable system, Theory
and practocal application, Wiley \& Son, Inc., 2006.

\bibitem{pa1} MacWilliams F. J., Sloane N. J.: The Theory of error-correcting codes, North-Holland Publishing
Company, Amsterdam, 1978.

\bibitem{pa2} \"{O}zen M., Siap V.: The MacWilliams identity for $m$-spotty weight enumerators of linear codes over finite
fields, Computers and Mathematics with Applications, Vol.61, No. 4,
1000-1004, (2011).

\bibitem{pa3} \"{O}zen M., Siap V.: The MacWilliams identity for $m$-spotty Rosenbloom-Tsfasman weight enumerator,
Journal of the Franklin Institute, http://dx. doi.org/10. 1016/j.
jfranklin. 2012. 06. 002,(2012).

\bibitem{pa4} Sharma A., Sharma A. K.: Sharma, On some new $m$-spotty Lee weight
enumerators, Des., Codes Cryptogr., Doi 10. 1007/s10623-012-9725-z,
(2012).

\bibitem{pa5} Sharma A., Sharma A. K.: On MacWilliams type identities for
$r$-fold joint $m$-spotty weight enumerators, Discret. Math.,
Vol.312, No. 22, 3316-3327, 2012.

\bibitem{pa6} Sharma A., Sharma A. K.: MacWilliams type identities for some new $m$-spotty weight enumerators,
IEEE Trans. Inform. Theory, Vol. 58, No. 6, 3912-3924, (2012).


\bibitem{pa7} Shi M. j.: The MacWilliams identity for $m$-spotty Rosenbloom-Tsfasman weight enumerator over finite
fields and ring $Z_l$, communicated for publication.

\bibitem{pa8} Siap I.: An identity between the $m$-spotty weight enumerators of a linear code and its dual, Turkish Journal
of Mathematics, doi:10.3906/mat-1103-55, (2012).

\bibitem{pa9} Siap I.: MacWilliams identity for $m$-spotty Lee weight enumerators, Appl. Math. Lett., Vol.23, Issune 1, 13-16, (2010).


\bibitem{pa10} Suzuki K., Kashiyama T., Fujiwara E.: MacWilliams identity for $m$-spotty weight enumerator, ISIT 2007, Nice, France, 31-35, (2007).

\bibitem{pa11} Suzuki K., Kashiyama T., Fujiwara E.: A general class of $m$-spotty weight enumerator,
IEICE-Trans. Fundam., Vol. E90-A, No.7, 1418-1427, (2007).

\bibitem{pa12} Umanesan G., Fujiwara E.: A class of random multiple bits in a byte error correcting and single byte error
detecting (St/bEC-SbED) codes, IEEE Trans. on Comput., Vol. 52,
No.7, 835-847, (2003).

\end{thebibliography}
\end{document}